\RequirePackage{fix-cm}

\documentclass{svjour3}                     
\smartqed  
\usepackage{graphicx}
\usepackage{physics}
\usepackage{dcolumn}
\usepackage{bm}
\usepackage{lipsum}
\usepackage{placeins}
\usepackage{mathtools}
\usepackage{cuted}
\usepackage{ amssymb }

\begin{document}

\title{A Novel Metal Nanoparticles- Graphene Nanodisks- Quantum Dots Hybrid-System-Based Spaser
}

\titlerunning{A Novel MNPs-GNDs-QDs hybrid system based spaser}        



\author{Mariam Tohari  \and Andreas Lyras \and Mohamad AlSalhi
}


\institute{Mariam Tohari  \at Department of Physics, College of Science, King Khalid University, Abha, P.O. Box 9004, Saudi Arabia\\
            \email{mrohary@kku.edu.sa}\and Andreas Lyras\at Department of Physics and Astronomy, College of Science, King Saud University, Riyadh, P. O. Box 11451, Saudi Arabia\\ \and Mohamad AlSalhi \at  Department of Physics and Astronomy, College of Science, King Saud University, Riyadh, P. O. Box 11451, Saudi Arabia\\
 Research Chair on Laser Diagnosis of Cancers, College of Science, King Saud University, Saudi Arabia.}           

\date{Received: date / Accepted: date}

\maketitle

\begin{abstract}
Active nanoplasmonics has recently led to the emergence of many promising applications. One of them is spaser (surface plasmons amplification by stimulated emission of radiation) that has been shown to generate coherent and intense fields of selected surface plasmon modes that are strongly localized in the nanoscale. We propose a novel nanospaser composed of metal nanoparticles-graphene nanodisks hybrid plasmonic system as its resonator and a quantum dots cascade stack as its gain medium. We derive the plasmonic fields induced by pulsed excitation through the use of the effective medium theory. Based on the density matrix approach and by solving the Lindblad quantum master equation, we get ultrafast dynamics of the spaser associated with coherent amplified plasmonic fields. The intensity of the latter is significantly affected by the width of the metallic contact and the time duration of the laser pulse used to launch the surface plasmons. The proposed nanospaser operates in the mid-infrared region that has received much attention due to its wide biomedical, chemical and telecommunication applications.
\end{abstract}

\section{Introduction}
\label{intro}

Over the last decade, plasmonic nanosources have attracted significant attention as coherent and intense near-field generators of nanolocalized optical fields, eliminating the need of a coupling mechanism between photons and surface plasmons (SPs) and paving the way to many promising applications \cite{gwo2016semiconductor,stockman2011nanoplasmonics}. The surface plasmon waves resulting from collective oscillations of free electrons near the surface of metals, propagate along the interface between metal and dielectric and decay in the direction perpendicular to the interface. The propagation length of surface plasmon polaritons is limited by the plasmonic losses that depend on the dielectric properties of metals \cite{gaponenko2010introduction}. For a metal bounded by an ideal dielectric, the losses are caused by free electron scattering and absorption through interband transitions at significantly short wavelengths \cite{khurgin2012reflecting}. 

Because the losses are often a serious limitation for practical applications of nanoplasmonics, many efforts have been devoted to proposing loss compensation and amplification approaches by introducing a gain medium in the dielectric surrounding the metal. Specifically, Seidel et al. reported the first experiment demonstrating amplification of surface plasmons on a flat silver film surrounded by dye molecules \cite{seidel2005stimulated}. This work was followed by many theoretical and experimental investigations \cite{ambati2008observation,de2008theory,ambati2008active,noginov2008compensation,oulton2009plasmon} that have paved the way to surface plasmon amplification by stimulated emission of radiation i.e., spaser, introduced by D. Bergman and M. Stockman \cite{bergman2003surface} and demonstrated experimentally with various plasmonic resonators, gain media and geometries \cite{de2010amplification,flynn2011room,de2011measuring,khurgin2012practicality,meng2013wavelength}. Moreover, several theoretical approaches have been established to explain the spasing quantum mechanically using two-level and three-level models for the gain medium \cite{stockman2010spaser,dorfman2013quantum}. Richter et al. have constructed a numerical approach to handle a large number of identical chromophores with an arbitrary number of energy levels non-perturbatively based on the density matrix approach using the Tavis-Cummings model in the Lindblad quantum master equation \cite{richter2015numerically}. 

Due to their ability to confine the optical energy near the surface, the spaser can utilize plasmonic components as its resonators to support the plasmonic modes and externally excited population-inverted gain media to provide the energy for spasing modes. The demonstration of spasing involves a resonant energy transfer from optical transitions in the gain medium to plasmon excitations in the metal, as well as stimulated emission of surface plasmons due to the high local fields created by plasmons that excite the gain medium and stimulate more emission of selected plasmonic modes leading to the required amplification \cite{stockman2008spasers}.     

The performance of a spaser is limited by the relaxation rates of the surface plasmons and the gain medium. To overcome the former limitation, plasmons in graphene provide a suitable alternative to those of metals due to the high mobility of its charge carriers leading to tight confinement and relatively long propagation distance as well as the tunability of graphene's plasmons via electrostatic gating \cite{falkovsky2007space,falkovsky2008optical,bolotin2008ultrahigh,koppens2011graphene}. A spaser formed by doped graphene nanoribbons surrounded by semiconductor QDs has been theoretically proposed \cite{berman2013graphene}. This spaser has been shown to support a wide frequency generation region from terahertz to infrared, small plasmon damping and low pumping threshold. Moreover, a tunable spherical graphene spaser that supports localized surface plasmon modes has been proposed. It was found that the spasing could occur when the quality factors of some localized modes becomes larger than some critical values given in terms of the Fermi energy of graphene \cite{ardakani2017tunable}. Additionally, V. Apalkov et al. have proposed a novel nanospaser made of a graphene nanopatch and a quantum well cascade emitter \cite{apalkov2014proposed}. With this spaser, optical fields have been generated exhibiting a high nanolocalization and coherent generation of SPs in the graphene nanopatch. However, the quantum well cascade suffers from nonradiative relaxation of electrons in the upper radiative state due to thermally activated electron-longitudinal optical phonon scattering \cite{li2013intersubband}. Interestingly, due to the discrete nature of their energy levels, it is possible to greatly increase the lifetime of the upper levels of QDs via phonon bottleneck that suppresses electron-longitudinal optical phonon scattering \cite{zibik2009long}. Thus, a spaser that utilizes quantum dots cascade emitter as its gain medium could be characterized by low spasing threshold \cite{michael2016mid}. Moreover, based on selection rules, the optical transitions in QDs cascade emitters are only allowed along the direction of the QDs growth \cite{paiella2006intersubband}. This can significantly simplify the calculations of a spaser. 

Recently, it has been shown that a metal nanoparticles- graphene nanodisks- quantum dots hybrid system can support ultrafast energy transfer between excitons and plasmons \cite{tohari2018ultrafast}. Specifically, within the near field approximation, having a relatively large size of metal nanoparticles (MNPs) of polarizability comparable to that of highly doped graphene nanodisks (GNDs) that support plasmons that are resonant with excitons in the QD, leads to a controllable and ultrafast energy transfer within the system \cite{tohari2018ultrafast}. Therefore, it is expected that by using a MNP-GND hybrid system as a resonator, one can enhance the performance of a spaser and exercise more control on its performance characteristics since the plasmons in graphene can be launched and controlled effectively with resonant metal nanoantennas \cite{alonso2014controlling,tohari2018ultrafast,tohari2018giant}. 

In the present work, we propose a novel nanospaser composed of a MNP-GND hybrid plasmonic system as its resonator and a quantum dots cascade as its gain medium to take advantage of the tunability of long-lived plasmons in graphene \cite{koppens2011graphene} and the possibility to control the latter by plasmons in noble metals, as well as the low spasing threshold  with a quantum dots cascade emitter. The proposed spaser operates in the mid-infrared region that has wide chemical, biomedical and telecommunication applications \cite{perenzoni2014physics}. The properties of the spaser as a plasmonic amplifier will be investigated using the density matrix theory with quantized plasmonic field in the transient regime.

 \section{Theoretical Fourmalim }
\label{theo}

\begin{figure}
\centering
\includegraphics[width=8 cm]{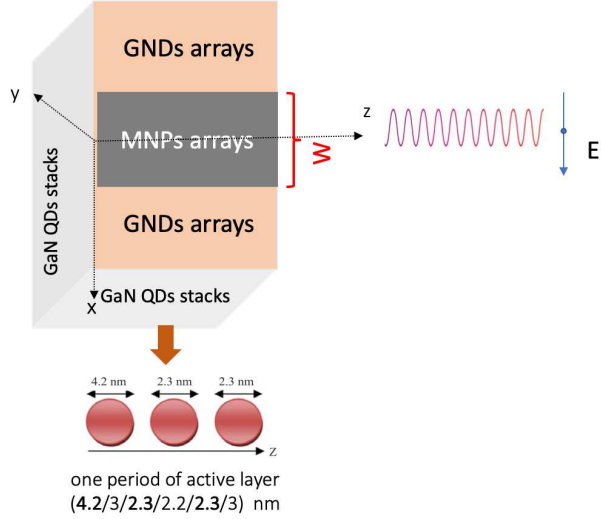}
\caption{The schematic illustration of the proposed MNP-GND-QD based spaser.}
\label{fig1}       
\end{figure}

We aim to study the properties of the proposed spaser depicted in Fig.~\ref{fig1} that utilizes the MNP-GND hybrid system as its resonator and GaN QDs cascade stack as its gain medium. The latter is characterized by mechanical and thermal stability as well as low sensitivity to ionizing radiation \cite{feng2006iii}. Moreover, GaN QDs cascade terahertz emitter, modeled for one period shown in Fig.~\ref{fig1}, where the bold numbers denote the size of the QDs that are sandwiched between $Al_{0.18}Ga_{0.82}N$ layers, have demonstrated a relatively large lifetime of the upper level leading to enhanced population inversion that supplies the spasing mode \cite{asgari2013modelling}. Let's first find the electric and magnetic field components of the plasmonic excitations near GNDs arrays taking into account the presence of the MNPs lattice. To this end, consider a two-dimensional periodic lattice of silver nanospheres of  width W, on top of a periodic lattice of GNDs, located at $z=0$, encompassing the $|x|<W/2$ region and centered at $x=0$.
   
Consider a transform limited pulse propagating along the z-direction and having its linear polarization along the interface between the GNDs layer and the dielectric (i.e. along the x-axis). The launching pulse induces the fields $E^{ind}$ and $B^{ind}$ that obey the Maxwell's equations and due to the symmetry in the y-direction, the fields have TM polarization with $E_{x}$, $E_{z}$ and $B_{y}$ nonzero field components.Thus by using Fourier transformation, one can write the field induced along the x direction as:   
\begin{equation}
E_{x}^{ind}(x,z,t)=\int_{-\infty}^{\infty}  e^{iqx-i\omega t} E_{x}^{ind}(q,z) dq
\label{eq1}
\end{equation}
 where $q$ is the wavenumber of the SPs waves propagating along the interface. In the following, the dependence on time will be dropped in order to simplify the notation and focus on the spatial dependence. Using the Maxwell's equations in the spectral representation, one can find:
 \begin{equation}
E_{x}^{ind}(q,z)=-i \frac{k_{q}}{2\epsilon_{0}\omega \epsilon\tilde{E}(\omega)}J_{x}(q) e^{-k_{q}|z|}
\label{eq2}
\end{equation}
where $k_{q}^{2}=q^2-\epsilon \frac{\omega^2}{c^2}$, and $\epsilon$ is the dielectric constant of the embedding medium. Note that $\tilde{E}(\omega)$ represents the  envelope of the pulse in its angular frequency domain. $J_{x}(q)$ is the Fourier component of the current density in the plane $z=0$ which can be written as:
 \begin{equation}
J_{x}(x)=\sigma_{GNDs}E_{x}(x,0)+(\sigma_{MNPs}-\sigma_{GNDs}) \Theta\left(\frac{W}{2}-|x|\right) E_{x}(x,0)
\label{eq3}
\end{equation}
where $\Theta$ is the Heaviside step function and $E(x,0)$ is the total electric field at $z=0$ given as the sum of the external and induced fields. $\sigma_{GNDs}$ and  $\sigma_{MNPs}$ are the effective conductivity of GNDs and MNPs arrays respectively. To model the optical response of nanoparticles arrays, it is fairly convenient to employ the effective medium approximation (EMA) that is only valid when the inner structure length scale is much smaller than the incident wavelength. Within EMA, the system comprising of interacting nanoparticles (NPs) is replaced by a homogeneous layer with an effective dielectric constant representing the overall response of the modeled system \cite{park2014optical}. The effective dielectric constant for nanodisks and nanospheres arrays has been derived analytically by Genov et al. through considering the RLC model of NPs arrays, in which the negative permittivity of plasmonic material and positive permittivity of dielectric are represented by inductance R-L and capacitance C respectively \cite{genov2004resonant}. Based on this model, for a plasmonic material, graphene (G) or metal (M), of dielectric constant written as $\epsilon_{M,G}=\epsilon_{M,G}^{'}(1-i\kappa_{M,G})$, with $\kappa_{M,G}<<1$, where $\kappa_{M,G}$ represents the loss parameter that is given as the ratio between the relaxation rate of plasmons $\gamma_{M,G}$ and the incident frequency $\omega$, the effective dielectric constant of periodic arrays of GNDs and metal nanospheres embedded in dielectric of $\epsilon_{d}$ are given as:
 \begin{subequations}
\label{eq4}
\begin{eqnarray}
\epsilon_{GNDs}=\epsilon_{G}^{'}\left[\frac{\pi}{\sqrt{2\kappa_{G}(\Delta_{G}-i)(P_{G}+1)}}-\frac{1+\pi/2}{(P_{G}+1)}\right]\label{4a}
\end{eqnarray}
\begin{equation}
\epsilon_{MNPs}=2\epsilon_{d}\left[(1+\frac{\kappa_{M}\Delta_{M}}{(P_{M}+1)})log\left(\frac{(P_{M}+1}{\kappa_{M}(\Delta_{M}-i)}\right)-1\right]\label{4b}
\end{equation}
\end{subequations}
where $P_{M,G}=|\epsilon_{M,G}^{'}|/\epsilon_{d}$, $\Delta_{M,G}=\left(\frac{P_{M,G}}{\delta_{M,G}}-1\right)/\kappa_{M,G}$. $\delta_{M,G}$ is the packing density of arrays given as a ratio between the diameter of particles and the interparticle distances. With plasma frequency $(\omega_{p})_{M,G}>>\omega$, $\epsilon_{M,G}^{'}$ can be approximated as $(\omega_{p}/\omega)_{M,G}^{2}$.

Substituting from Eq.~\ref{eq2} into Eq.~\ref{eq1}, through the use of Fourier transformation of $J_{x}(x)$ leads to \cite{satou2007excitation}:
 \begin{equation}
E_{x}^{ind}(x,z)=-\frac{i}{4\pi \epsilon_{0}\epsilon \omega \tilde{E}(\omega)}\int_{-\infty}^{\infty}dq k_{q}e^{iqx-k_{q}|z|}\int_{-\infty}^{\infty} dx^{'}J_{x}(x^{'})e^{-iqx^{'}}
\label{eq5}
\end{equation}
Combining Eqs.~\ref{eq5} and ~\ref{eq3} leads to the following equation for  $E_{x}(x,0)$ in all space \cite{satou2007excitation}:
  \begin{eqnarray}
E_{x}(x,0)=&&\frac{1}{2\pi \tilde{E}(\omega)}\left( 1-\frac{\sigma_{GNDs}}{\sigma_{MNPs}} \right)\int_{-\infty}^{\infty}dq\frac{e^{iqx}}{\xi_{q}}\int_{-W/2}^{W/2}dx^{'}e^{-iqx^{'}}E_{x}(x^{'},0)\nonumber\\&&+\frac{\sigma_{GNDs}}{\sigma_{MNPs}}\frac{E_{0}}{\tilde{E}(\omega)\xi_{0}}
\label{eq6}
\end{eqnarray}
  where $\xi_{q}=1+(i\sigma_{GNDs}k_{q}/2\omega \epsilon_{0}\epsilon)$ is the dielectric function of the GNDs layer. Clearly, $E_{x}(x^{'},0)$ involved in Eq.~\ref{eq6}, represents the field inside the metallic contact of width $W$ that can be written as a Fourier expansion:
    \begin{equation}
E_{x}(x^{'},0)=\sum_{n=0}^{\infty}A_{n}cos\left(  \frac{2\pi n x^{'}}{W} \right)
\label{eq7}
\end{equation}
Thus, Eq.~\ref{eq6} can be reduced to a system of algebraic equations to be solved numerically in a matrix form running the sum up to a judicially chosen upper value $n=N$. After determining the coefficients $A_{n}$, the electric field induced along the x direction near graphene layer due to the inhomogeneity introduced by the metallic contact is obtained in terms of the dimensionless quantities; $u=qW$ and $k_{u}=k_{q}W$ as:
 \begin{equation}
 E_{x,u}(x,z)=\frac{2\zeta_{\sigma}}{\tilde{E}(\omega)\pi}\sum_{n=0}^{N}(-1)^{n}A_{n}\int_{0}^{\infty}du\frac{1-\xi_{u}}{\xi_{u}}\frac{u sin (u/2)}{u^{2}-4n^{2}\pi^2}cos(ux/W)e^{-k_{u}|z|/W}
 \label{eq8}
\end{equation}
where $\zeta_{\sigma}=(\sigma_{MNPs}/\sigma_{GNDs})-1$ and
 \begin{equation}
\xi_{q}=1+\frac{i\sigma_{GNDs}}{2\omega \epsilon_{0}\epsilon_{2} W}\sqrt{u^{2}-a^{2}} \quad \textrm{and} \quad a^{2}=\frac{\epsilon_{2}\omega^{2}W^{2}}{c^{2}}
 \label{eq9}
\end{equation}
It is clear that the parameter $a$ contains a dependence on the ratio of the width of metallic contact to the incident wavelength. Thus, $a$ can be considered as a measure of raterdation effects. Specifically, the case of small $a $ i.e, $u>a$ corresponds to evanescent wave confined near graphene layer. Eq.~\ref{eq8} gives the electric field component of the induced field along the interface. The transverse component can be found by using the following Maxwell's equations valid for TM polarization:
\begin{subequations}
\label{eq10}
\begin{eqnarray}
E_{z}^{ind}(x,z)=i\frac{c^{2}}{\omega \epsilon}\partial_{x}B_{y}^{ind}(x,z)
\label{eq10a}
\end{eqnarray}
\begin{equation}
B_{y}^{ind}(q,z)=i\frac{\omega \epsilon}{c^{2}k_{q}^{2}}\partial_{z}E_{x}^{ind}(q,z)
\label{eq10b}
\end{equation}
\end{subequations}
 Using Eq.~(\ref{eq2}), and substituting from Eq.~(\ref{eq10b}) into Eq.~(\ref{eq10a}) with inserting $J_{x}$ given through  Eq.~(\ref{eq3}), we obtain the two-dimensional electric field induced along $z$ direction due to the metallic contact:
 \begin{eqnarray}
 E_{z,u}(x,z)=&&-sgn(z)\frac{2\zeta_{\sigma}}{\tilde{E}(\omega)\pi}\sum_{n=0}^{N}(-1)^{n}A_{n}\int_{0}^{\infty}du\frac{1-\xi_{u}}{\xi_{u}}\frac{u^{2} sin (u/2)}{u^{2}-4n^{2}\pi^2}\frac{sin (ux/W)}{k_{u}}\nonumber\\&& \times e^{-k_{u}|z|/W}
 \label{eq11}
\end{eqnarray}
 The corresponding transverse component of the magnetic field can be obtained using Eq.~(\ref{eq10a}):
 \begin{eqnarray}
 B_{y,u}(x,z)=&&sgn(z)\frac{2iW\omega\epsilon}{c^{2}}\frac{\zeta_{\sigma}}{\tilde{E}(\omega)\pi}\nonumber\\&&\sum_{n=0}^{N}(-1)^{n}A_{n}\int_{0}^{\infty}du\frac{1-\xi_{u}}{\xi_{u}}\frac{u sin (u/2)}{u^{2}-4n^{2}\pi^2}\frac{cos (ux/W)}{k_{u}}e^{-k_{u}|z|/W}
 \label{eq12}
\end{eqnarray}
The propagation of surface plasmons is governed by the dispersion relation derived by using the condition of the discontinuity of $B_{y}$ at $z=0$. For a relatively large $\sigma_{GNDs}$, i.e. $\frac{1-\xi_{u}}{\xi_{u}}\approx-1$, the discontinuity of $B_{y}$ is reduced to the following dispersion relation:
\begin{equation}
u=-\frac{W\omega (\epsilon_{1}+\epsilon_{2})}{\sigma}
 \label{eq13}
\end{equation}
where $\sigma$ is the total conductivity of the GNDs and MNPs arrays at  $z=0$ and $\epsilon_{1}$ ($\epsilon_{2}$) is the electric permittivitiy of the dielectric medium in $z>0$ ($z<0$). Note that the frequency of the surface plasmon polaritons for the system can be determined by solving for the pole of the dispersion relation that is given in terms of geometrical features and optical properties of the system.

    To study the properties of the proposed MNPs-GNDs-QDs hybrid-system-based spaser, we focus on the z-component of the electric field, that describes the confined plasmonic field in the direction perpendicular to the interface, since the dipole transitions of the gain medium is only allowed along this direction \cite{paiella2006intersubband}. Let the wavenumber-dependent amplitude in the field equation be:
   \begin{equation}
\sum_{n=0}^{\infty}A_{n}\frac{usin(u/2)}{u^{2}-4n^{2}\pi^{2}}=\varepsilon_{u}
 \label{eq14}
\end{equation}
 $\varepsilon_{u}$ can also be written in terms of surface plasmon polaritons energy $\hbar \omega_{u}$ where the field can be quantized by using the Brillouin expression for the field mean energy in the dispersive medium \cite{landau2013electrodynamics,apalkov2014proposed}:

\begin{equation}
\frac{1}{4}\left[ \int d^{3} \bm{r} \frac{d(\omega \epsilon(\omega,\bm{r}))}{d\omega}\left|\bm{E}_{u}(\bm{r}) \right|^{2}+ \int d^{3} \bm{r}\frac{ B_{uy}^{2}(\bm{r})}{\mu^{2}} \right]=\hbar \omega_{u}
 \label{eq15}
\end{equation}
   where $\mu$ and $\epsilon(\omega,\bm{r})$ are magnetic permeability and  electric permittivity of the dispersive medium respectively. The latter can be defined as:
   \begin{equation}
\epsilon(\omega,\bm{r})=\frac{\sigma(\omega)}{\omega}\delta(z)+\epsilon_{1}\Theta(z)+\epsilon_{2}(1-\Theta(z))
 \label{eq16}
\end{equation}
Solving Eq.~\ref{eq15} yields the z-component of the quantized plasmonic field of mode $\hbar \omega_{u}$ in the $z<0$ region of the dielectric medium ( medium 2) where the gain medium is placed:
  \begin{equation}
\hat{E}_{z,u}=\sqrt{\frac{4\hbar \omega_{u} u}{S_{0}W(\epsilon_{1}+\epsilon_{2})\left[-\frac{5}{4}-\frac{2\omega_{u}}{u} \frac{du}{d\omega_{u}}\right]}} sin (ux/W)e^{-k_{2u}|z|/W}(\hat{a}_{u}^{\dagger}+\hat{a}_{u})
 \label{eq17}
\end{equation}
where $\hat{a}_{u}^{\dagger}$ and $\hat{a}_{u}$ are raising and lowering operators of SPs respectively. $|z|$ represents the penetration length of plasmonic field that can be adjusted to be the distance between the GNDs layer and the active medium. The area of the GNDs layer is denoted by $S_{0}$ and $du/d\omega_{u}$ included in Eq.~\ref{eq17} can be obtained by using the dispersion relation, Eq.~\ref{eq13}:
\begin{eqnarray}
\frac{du}{d\omega_{u}}=&&-(\epsilon_{1}+\epsilon_{2})W \nonumber\\&& \left[ \frac{2\sigma_{GNDs}+[\sigma_{MNPs}-\sigma_{GNDs}]\frac{W}{L}}{\left(2\sigma_{GNDs}+[\sigma_{MNPs}-\sigma_{GNDs}]\frac{W}{L}\right)^{2}}\right]\nonumber\\&&-(\epsilon_{1}+\epsilon_{2})W\left[ \frac{-\omega_{u}\left(2E(\omega_{u})\right)+F(\omega_{u})\frac{W}{L}-2E(\omega_{u})\frac{W}{L}}{\left(2\sigma_{GNDs}+[\sigma_{MNPs}-\sigma_{GNDs}]\frac{W}{L}\right)^{2}}\right]
 \label{eq18}
\end{eqnarray}
where $E(\omega_{u})$ and $F(\omega_{u})$ are defined in Appendix A. 

\begin{figure}
\centering
\includegraphics[width=10cm]{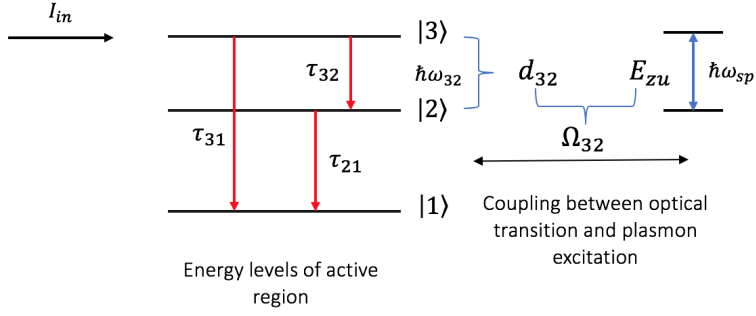}
\caption{The model of the proposed MNPs-GNDs-QDs hybrid-system-based spaser.}
\label{fig2}       
\end{figure}

The model of the proposed mid-infrared MNPs-GNDs-QDs hybrid-system-based spaser is shown in Fig.~\ref{fig2}. Following injection current pumping $I_{in}$, the electrons injected into the upper subband level of the GaN QDs cascade emitter, undergo stimulated emission to the lower level (level2). Subsequently, these electrons relax nonradiatively to the lowest energy level (level 1) to eventually tunnel via the chirped superlattice to the adjacent QD  with rate $\gamma_{out}$ \cite{paiella2006intersubband}.  

In the presence of a resonant SPs mode with the dipole transition energy $\hbar \omega_{32}$ of the GaN QDs cascade gain medium, energy transfer will occur through the coupling between the dipole transitions of the gain medium and plasmon excitations leading to amplification of this mode of SPs. Note that, the number of periods of the QDs cascade emitter that can be included in the energy transfer depends on the penetration length of the plasmonic fields. 

The dynamics of the system can be investigated by solving the Lindblad master equation through the use of the following Hamiltonian for the $u$ mode:
\begin{equation}
H_{u}=\hbar \omega_{u} \hat{a}_{u}^{\dagger}\hat{a}_{u}+\frac{1}{2}\hbar\omega_{32}(\sigma_{33}-\sigma_{22})-\hbar \left[ \Omega_{32}\ket{3}\bra{2} \hat{a}_{u}+\Omega_{32}^{*} \ket{2}\bra{3}\hat{a}_{u}^{\dagger} \right]
 \label{eq18}
\end{equation}
where $\Omega_{32}$ is the coupling parameter illustrated in Fig.~\ref{fig2} and defined as $d_{32}E_{uz}/\hbar$. $d_{32}$ is the dipole moment of transition $\ket{3} \leftrightarrow\ket{2}$. Since the dipole transition of the GaN QDs cascade is along the z-direction,  we consider only the z-component of the plasmonic fields. The two first terms of the Hamiltonian refer to the plasmonic resonator with mode $u$ and the gain medium respectively. The third term describes the energy transfer between the optical transitions in the gain medium and plasmon excitations. Specifically, $\ket{3}\bra{2} \hat{a}_{u}$ corresponds to the absorption of a SP in order to make the resonant transition from $\ket{2}$ to$\bra{3}$. On the other hand, $ \ket{3}\bra{2}\hat{a}_{u}^{\dagger} $ represents the emission of a SP to make the transition from $\ket{3}$ to$\bra{2}$. The Liouvillian of the system that describes the decay channels is given by \cite{scully1997quantum}:
\begin{eqnarray}
{\mathcal{L}_{\rho}}=&&\frac{\gamma_{31}}{2}(\rho \sigma_{33}+\sigma_{33} \rho-2\sigma_{13}\rho\sigma_{31})\nonumber\\
&& +\frac{\gamma_{32}}{2}(\rho \sigma_{33}+\sigma_{33} \rho-2\sigma_{23}\rho\sigma_{23})\nonumber\\
&& +\frac{\gamma_{21}}{2}(\rho \sigma_{22}+\sigma_{22} \rho-2\sigma_{12}\rho\sigma_{12})\nonumber\\&&  +\gamma_{out}(\rho \sigma_{11}+\sigma_{11}\rho)
\label{eq19}
\end{eqnarray}
where $\gamma_{out}$ is the rate of tunneling given in terms of the dimension of QD ($L_{QD}$), and its effective mass ($m_{QD}$) as \cite{asgari2013modelling} :

\begin{equation}
\gamma_{out}=\frac{\pi \hbar}{(2L_{QD}^{2}m_{QD})exp\left[\frac{2L_{b}\sqrt{2m_{b}H_{b}}}{\hbar}\right]}
 \label{eq20}
\end{equation}
$L_{b}$ and $m_{b}$ in Eq.~\ref{eq20} are the thickness of barrier and the effective mass respectively. The height of the potential barrier is denoted by $H_{b}$. With the above considerations, the density matrix equations of the system through the use of the Lindblad master equation are given as:
\begin{subequations}
\label{eq21}
\begin{equation}
\dot{\rho}_{032}=-\left[\left(\frac{\gamma_{31}}{2}+\frac{\gamma_{32}}{2}+\frac{\gamma_{21}}{2}\right)+i(\omega_{32}-\omega_{sp})\right]\rho_{032}+i\Omega_{32}a_{0u}(\rho_{33}-\rho_{22})+\frac{I_{in}}{e},
\label{eq21a}
\end{equation}
\begin{equation}
\dot{\rho}_{031}=-\left(\frac{\gamma_{31}}{2}+\frac{\gamma_{32}}{2}\right)\rho_{031}-i\left[ \frac{\omega_{32}}{2}-(\omega_{sp}-i\gamma_{21}) \right]\rho_{031}+i\Omega_{32}\rho_{021}a_{0u}+\frac{I_{in}}{e},
\label{eq21b}
\end{equation}
\begin{equation}
\dot{\rho}_{021}=i\left[(\frac{\omega_{32}}{2}-i\gamma_{21})+i\gamma_{out} \right]\rho_{021}+i\Omega_{32}^{*}a_{0u}^{*}\rho_{031},
\label{eq21c}
\end{equation}
\begin{equation}
\dot{\rho}_{33}=-(\gamma_{31}+\gamma_{32})\rho_{33}-\left(i\Omega_{32} \rho_{023}a_{0u}-i\Omega_{32}^{*} \rho_{032}a_{0u}^{*}\right)+\frac{I_{in}}{e},
\label{eq21d}
\end{equation}
\begin{equation}
\dot{\rho}_{22}=-\gamma_{32}\rho_{33}+\gamma_{21}\rho_{22}+\left(i\Omega_{32} \rho_{023}a_{0u}-i\Omega_{32}^{*} \rho_{032}a_{0u}^{*}\right),
\label{eq21e}
\end{equation}
\begin{equation}
\dot{\rho}_{11}=\gamma_{31}\rho_{33}+\gamma_{21}\rho_{22}-\gamma_{out}\rho_{11},
\label{eq21f}
\end{equation}
\begin{equation}
\dot{a}_{0u}=\left[ i(\omega_{32}-\omega_{sp})-\gamma_{sp}\right]a_{0u}-i\rho_{032}\Omega_{32}^{*} ,
\label{eq21g}
\end{equation}
\end{subequations}
where $a_{u}=a_{0u}e^{-i\omega_{sp}t}$, $\rho_{32}=\rho_{032}e^{-i\omega_{sp}t}$, $\rho_{31}=\rho_{031}e^{-i(\omega_{sp}-i\gamma_{21})t}$ and $\rho_{21}=\rho_{021}e^{-\gamma_{21}t}$, with slowly varying amplitudes; $a_{0u}$, $\rho_{032}$, $\rho_{031}$ and $\rho_{021}$. $\gamma_{sp}$ is the damping rate of SPs. Note the rate of injection current i.e, $I_{in}/e$, where $e$ is charge of electron, affects the coherence and population terms that are related to the upper level \cite{paiella2006intersubband}. $\gamma_{ij}$ are the damping rates corresponding to the relaxation times ($\tau_{ij}$) shown in Fig. ~\ref{fig2}.

\section{Analysis of the confined plasmonic fields near the GNDs layer}
\label{sec:2}

\begin{figure}
\centering
\includegraphics[width=12cm]{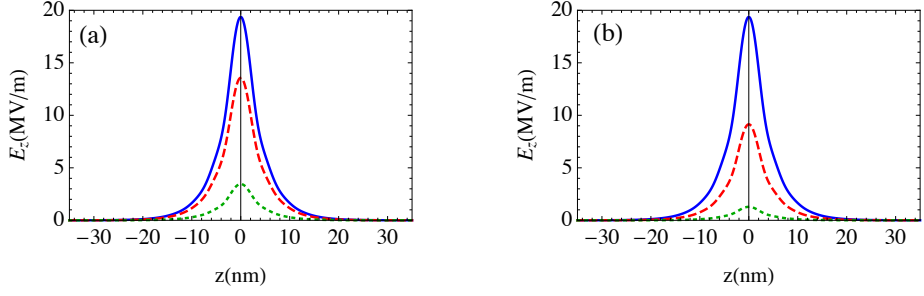}
\caption{The confined plasmonic fields near the GNDs layer with thickness of $t=0.34nm$, Fermi energy of $E_{F}=1eV$, mobility of $20000cm^{2}/Vs$ and $\delta_{GNDs}=100$ excited by a $10fs$-sech pulse of wavelength $\lambda_{0}=30\mu m$, in the presence of the metallic contact i.e., Ag NPs lattice of $\delta_{AgNPs}=150$ and $(\bm{a})$ width of; $W=0.25\lambda_{0}$ (solid),  $W=0.5\lambda_{0}$ (dashed) and $W=1.2\lambda_{0}$ (dotted) at $x/W=0.45$. The two lattices are embedded in a dielectric of $\epsilon=3$. $(\bm{b})$ width of $W=0.25\lambda_{0}$ for the metalic contact at $x/W=0.45$. The two lattices are embedded in dielectric medium of $\epsilon=3$ (solid), $\epsilon=6$ (dashed), $\epsilon=10$ (dotted).}
\label{fig3}       
\end{figure}

 This paper aims to study the optical properties of a spaser consisting of a GNDs lattice excited by means of a MNPs one, as its resonator and GaN QDs cascade emitter as its gain medium. Thus, since the latter has optical transitions only along the direction of the QDs growth, i.e. the z-direction, we focus on the z-component of electric field given by Eq.~\ref{eq11}, that describes the confined plasmonic fields near the GNDs layer. Fig.~\ref{fig3} shows the confined plasmonic field over a few tens of nanometers corresponding to the penetration length of the plasmonic field, where the sum in Eq.~ \ref{eq11} is running up to $N=20$. Clearly, with a 10 fs-sech launching pulse of time dependence given in terms of the time duration of the pulse $\Delta\tau$ as \cite{shapiro1984ultrashort}:
 
 \begin{equation}
E(t)=\left( \frac{log (1+\sqrt{2})}{\Delta\tau}  \right)^{1/2}sech\left( \frac{2t log (1+\sqrt{2})}{\Delta\tau}  \right)
\label{eq22}
\end{equation}     
 we get a relatively strong confined plasmonic fields of $MV/m$ strength that can be significantly enhanced with shorter pulsed excitation. Moreover, the confined plasmonic fields are enhanced for a relatively small ratio between the width of the metallic contact and the wavelength of the incident radiation, as well as small dielectric constant of the medium where the two lattices are embedded. This can be attributed to the corresponding limited retardation effects that are responsible for the decay of the confined surface plasmons as an electromagnetic radiation \cite{mikhailov2005microwave}. 
 
\FloatBarrier
   
\begin{figure}
\centering
\includegraphics[width=6cm]{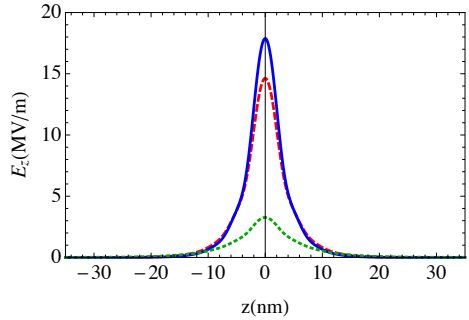}
\caption{The confined plasmonic fields near the GNDs layer with thickness of $t=0.34nm$, Fermi energy of $E_{F}=1eV$, mobility of $20000cm^{2}/Vs$ excited by a $10fs$-sech pulse of  $\lambda_{0}=30\mu m$, in the presence of metallic contact i.e., Ag NPs lattice of width  $W=0.25\lambda_{0}$ at $x/W=0.45$ with $\delta_{GNDs}=50$-$\delta_{AgNPs}=100$ (solid),  $\delta_{GNDs}=70$-$\delta_{AgNPs}=100$ (dashed) and  $\delta_{GNDs}=200$-$\delta_{AgNPs}=100$ (dotted). The two lattices are embedded in dielectric of $\epsilon=3$.}
\label{fig4}       
\end{figure}

  Additionally, we examine the effect of the packing density of the GNDs and Ag NPs lattices on the confinement of the plasmonic fields near the GNDs layer as illustrated in Fig.~\ref{fig4}. We observe that a relatively strong confined plasmonic field is obtained when the packing density of the Ag NPs lattice is larger than that of the GNDs one. The confined plasmonic field decreases as the difference of packing density between the two lattices decreases. We observe a significant decrease of the confined plasmonic field for the case of $\delta_{GNDs} > \delta_{AgNPs}$. This is reasonable since the plasmonic fields are enhanced for large $\delta_{AgNPs}/\delta_{GNDs}$ corresponding to a highly conductive stripe. Note that, the conductivities of the two lattices are increased for a large packing density \cite{genov2004resonant}.
  
\begin{figure}
\centering
\includegraphics[width=12cm]{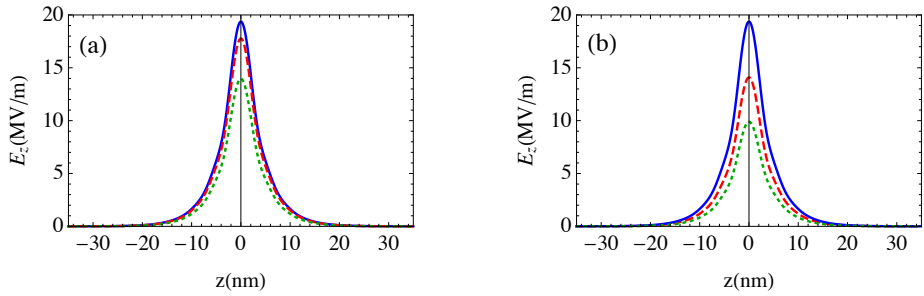}
\caption{The confined plasmonic fields near the GNDs layer with mobility of $20000cm^{2}/Vs$ excited by a $10fs$  sech pulse of  $\lambda_{0}=30\mu m$, in the presence of metallic contact i.e., Ag NPs lattice of width $W=0.25\lambda_{0}$ at $x/W=0.45$ with $\delta_{GNDs}=100$-$\delta_{AgNPs}=150$. The two lattices are embedded in dielectric of $\epsilon=3$. The thickness and doping level of GNDs are; $(\bm{a})$ $t=0.34nm$, $E_{F}=1eV$ (solid), $E_{F}=0.7eV$ (dashed) and $E_{F}=0.4eV$ (dotted). $(\bm{b})$ $E_{F}=1eV$, $t=0.34nm$ (solid), $t=0.55nm$ (dashed), $t=0.6nm$ (dotted).}
\label{fig5}       
\end{figure}

Fig.~\ref{fig5} shows the effect of the thickness of the GNDs $(t)$ and their doping levels, that are defined by the Fermi energy ($E_{F}$), on the confined plasmonic fields. Clearly, the latter are enhanced for a relatively small thickness of GNDs with large doping levels that lead to a relatively large bulk plasma frequency for graphene, $\omega_{p,G}^{2}=(e/\hbar)^{2}E_{F}/\pi \epsilon_{0}t$. It can be seen that the penetration length is not affected by the thickness of the GNDs and their doping levels. Additionally, the confined plasmonic field near the GNDs layer is sensitive to the thickness of the GNDs more than the doping level. 

  \section {Analysis of dynamics of the proposed MNPs-GNDs-QDs hybrid system-based spaser}
  
  Unlike the conventional lasers and amplifiers in quantum electronics, spasers have an inherent feedback provided by plasmonic structures that typically can not be removed \cite{stockman2011nanoplasmonics}. Thus, a spaser will develop the accumulation of a large number of coherent SPs leading to the CW regime where the gain compensates exactly for the losses with zero net amplification. One way to operate a spaser as a plasmonic amplifier, is to consider the transient regime recalling the fact that the establishment of  the CW regime requires a relatively long time \cite{stockman2010spaser}. Thus, we consider the transient amplification of SPs during hudreds of femtoseconds, a time interval smaller than the relaxation times of the gain medium and SPs. The transient dynamics of the spaser can be directly obtained by the numerical solution of the density matrix equations (Eqs.~\ref{eq21}).
  
\begin{figure}
\centering
\includegraphics[width=12cm]{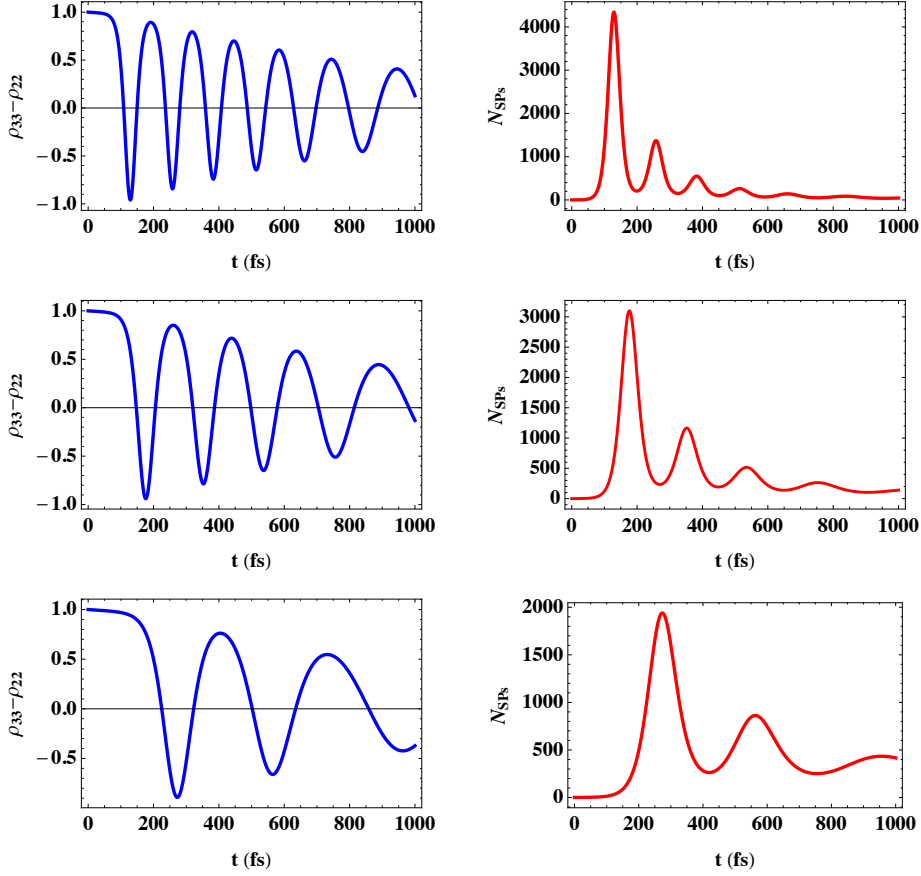}
\caption{Ultrafast dynamics of the proposed MNPs-GNDs-QDs hybrid-system-based spaser. ($\bm{a,c,e}$) The temporal behavior of the population inversion. ($\bm{b,d,f}$) The time evolution of SPs population in the spasing mode of energy $41.5meV$ with metallic width ($\bm{a,b}$): $W=0.2\lambda_{0}$,  ($\bm{c,d}$): $W=0.4\lambda_{0}$ and ($\bm{e,f}$): $W=0.6\lambda_{0}$. $\lambda_{0}$ is the wavelength of a $1ps$-pulse excitation.}   
\label{fig6}       
\end{figure}

At first, we solve numerically for the pole of the dispersion relation of SPs for the planar structure given by Eq.~\ref{eq13}. The parameters of the system are adjusted to get SPs mode energy resonant with that of the GaN QDs cascade emitter. In the particular, for a GNDs monolayer of  thickness  $0.34nm$, Fermi energy of $0.7eV$ and mobility of a $10^4cm^2/Vs$ with the GNDs lattice (Ag NPs lattice) embedded in a dielectric medium of dielectric constant 3 (12) with packing density 100 (150), we get SPs mode energy $41.5 meV$ resonant with the gain medium \cite{asgari2013modelling}. The relaxation rates of the gain medium are set to be $\gamma_{31}=7.14\times 10^{10}s^{-1}$, $\gamma_{32}=7.7\times 10^{10}s^{-1}$ and $\gamma_{21}=2.2\times 10^{12}s^{-1}$ \cite{asgari2013modelling} whereas that of SPs corresponding to $E_{F}=0.7eV$ is $\gamma_{sp}=7.12\times10^{11} s^{-1}$. The rate of injection is set to be $10^{10}s^{-1}$ and the rate of tunneling is calculated with GaN QD of dimension $L_{QD}=4nm$ and effective mass of $0.2m_{0}$ with AlGaN barrier of $L_{b}=3nm$ and $m_{b}=0.3m_{0}$ correspondingly
\cite{levinshtein2001properties}. The dipole moment of the gain medium is taken as $2.5 e$ $nm$ \cite{michael2016mid} and the area of the GNDs layer is set to be  $75\mu m^{2}$. The plasmonic components are launched by a 1ps-sech pulse to be shorter than the relaxation time of the gain medium where the slowly varying envelop approximation is valid \cite{boyd2003nonlinear} to simplfy our calculations for energy transfer between the plasmonic excitations and the optical transition of the gain medium. The dynamics of energy transfer between the SPs and the optical transitions of the gain medium are shown in Fig.~\ref{fig6} by numerically solving for the time-dependent density matrix equations (Eqs.~\ref{eq21}) with initial conditions; $\rho_{33}(0)=a_{0u}^{*}(0)=1$. The penetration length of the plasmonic field is set to be equivalent to one period of the GaN QDs cascade emitter i.e., $z=17nm$ \cite{asgari2013modelling}. It can be seen that transient amplification of SPs mode of energy $41.5meV$ is obtained over hundreds  of femtoseconds before they are damped depending on the relaxation rates of the SPs and the gain medium. It is remarkable that at the instant the number of SPs is maximized,  the population inversion reaches its minimum implying that the energy of optical transition is completely transferred to the SPs to be amplified. Interestingly, the amplification of SPs is significantly affected by the width of the metallic contact emphasizing the important role of the metal in the energy transfer as shown in Ref. \cite{tohari2018ultrafast}.  Specifically, for large width of the metallic contact, the amplification of SPs is decreased and the time corresponding to the maximum number of SPs is displaced to later values. This can be attributed to the retardation effects that dominate as the width of the metallic contact approaches the wavelength of the incident radiation leading to the decay of the SPs \cite{downing2017retardation}.

\FloatBarrier

\begin{figure}
\centering
\includegraphics[width=7cm]{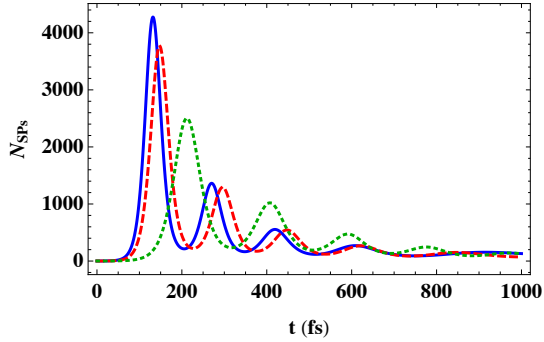}
\caption{Ultrafast dynamics of the SPs in the proposed MNPs-GNDs-QDs hybrid-system-based spaser with metallic contact of width $W=0.3 \lambda_{0}$, induced by a sech-pulse excitation of $\Delta\tau=700fs$ (solid), $\Delta\tau=900fs$ (dashed) and $\Delta\tau=2ps$ (dotted).}
\label{fig7}       
\end{figure}

Fig.~\ref{fig7} shows the effect of the time duration of the normalized sech-pulse used for launching SPs on the dynamics of the SPs mode energy $41.5meV$. It can be seen that the amplification of SPs is enhanced for small time duration of the launching pulse that is associated with a relatively large intensity. This enhancement is due to the large local field enhancement induced near of the surface of plasmonic materials with pulsed excitation \cite{villarreal2018surface}. Now, to check the degree of coherence of the amplified SPs mode of energy $41.5meV$, we calculate the second-order correlation function of the amplified plasmonic field that takes the form \cite{fox2006quantum}: 
\begin{equation}
g^{(2)}(\tau_{d})=\frac{\left<a_{u}^{\dagger}a_{u}^{\dagger} a_{u} a_{u}\right>}{\left<a_{u}^{\dagger}a_{u} \right>^{2}},\label{eq23}
\end{equation}   
where $\tau_{d}$ is the time delay. We calculate the time average over the period of the pulse oscillation since we are interested in the transient solution to obtain $g^{(2)}(\tau_{d}=0)=1$, implying that the amplified plasmonic field is perfectly coherent. This coherence can be attributed to the relatively low relaxation rate of the upper level of the spasing transition. In particular, since the transient amplification occurs over hundreds of femtoseconds a time duration smaller than the relaxation time of the upper level, the spontaneous emission noise can be avoided. Moreover, the spasing threshold of our proposed spaser consisting of GNDs-MNPs plasmonic system and GaN QDs cascade emitter that are separated by a distance $l$ can be easily estimated as a compersion between losses and gain, $\hbar^{2} |\Omega_{32}|^{2} e^{-2ql} \geq \Gamma_{SP} \Gamma_{32}$. Interestingly, the spasing threshold of the proposed spaser is smaller than that demonstrated for the graphene spaser proposed by Apalkov  et al.\cite{apalkov2014proposed} due to the relatively small intersubband-polarization relaxation width of the gain medium ($\Gamma_{32}=50\mu eV$) and the relatively low damping width of the SPs ($\Gamma_{SP}=470\mu eV$).

\FloatBarrier

Clearly, the transient regime represents the region where the spaser can act as a plasmonic amplifier with gain exceeding the losses achieving a coherent amplification of a selected mode of SPs that can be controlled by the width of metallic contact and the time duration of the launching pulse leading to enhancement of plasmonics.
\section{Conclusions}

In this paper, we have studied the properties of the proposed MNPs-GNDs-QDs hybrid-system-based spaser. Firstly, we investigated the confined plasmonic fields near the interface of the GNDs lattice with a metallic contact, represented by a Ag NPs lattice, induced by the launching pulse that has its linear polarization along the interface between the GNDs layer and the dielectric environment. We have found that the plasmonic field can be enhanced for a small width of the metallic contact and small dielectric constant of the surrounding medium where the retardation effects are limited. Moreover, the confined plasmonic field is enhanced for a small thickness of the GNDs and large doping levels that lead to a relatively large density of charge carriers. Large packing density in the  MNPs lattice compared to that of the GNDs can also significantly enhance the confined plasmonic fields. 

Subsequently, we have studied the dynamics of the proposed spaser with a GaN QDs cascade emitter chosen to be the gain medium. Ultrafast dynamics are obtained with transient amplification of SPs mode providing a plasmonic amplifier. The number of SPs that determines the intensity of the obtained plasmonic field is significantly affected by the width of the metallic contact emphasizing the important role of the MNPs lattice in the spasing process. In particular, the presence of MNPs lattice provides more options for controlling and enhancing the amplification of the plasmonic fields. Due to the relatively small spontaneous decay rate of the gain medium, perfectly coherent amplified plasmonic fields are obtained with a relatively low spasing threshold.

Interestingly, the factors that enhance the confined plasmonic fields act cooperatively and cumulatively to enhance the performance of the proposed plasmonic amplifier implying that these factors are absolutely realistic and can be accommodated by the present technology. Specifically, the obtained controllable and coherent SPs amplification could prove useful for various interesting applications  such as stand-alone components or gain sections integrated with other plasmonic elements to compensate for the losses and improve the performance of plasmonic devices.
 
 \section{Acknowledgements}

 The authors would like to express their gratitude to King Khalid University, Saudi Arabia for providing administrative and technical support.
 
\nocite{*}
\begin{appendix}
\section{Calculations of the frequency derivative of the GNDs and MNPs lattices' conductivities}
\unskip
\subsection{The frequency derivative of $\sigma_{GNDs}$}
The conductivity of the GNDs lattice can be given by \cite{genov2004resonant}:
\begin{equation}
\sigma_{GND}(\omega)=i\omega \epsilon_{0}t\left[ 1-|\epsilon^{'}(\omega)|\left( \frac{\pi}{\sqrt{2\frac{\gamma_{G}}{\omega}(\Delta_{G}-i)(P_{G}+1)}}-\frac{1+\pi/2}{P_{G}+1}\right) \right]
\label{A1}
\end{equation}
where:
\begin{equation}
|\epsilon^{'}(\omega)|=Re\left[1+i\frac{\sigma_{G}}{t\omega \epsilon_{0}}\right]
\label{A2}
\end{equation}
$\sigma_{G}$ is the Drude conductivity of graphene at low frequencies, $\gamma_{G}$ is the damping rate of graphene plasmons and $t$ is the thickness of graphene. $\Delta_{G}=\omega (P_{G}/\delta_{G}-1)/\gamma_{G}$ and $P_{G}=|\epsilon^{'}(\omega)|/\epsilon_{d}$. The packing density of the GNDs lattice is given by $\delta_{G}$. Using the Drude conductivity of graphene one can obtain:
\begin{equation}
 \left[ \frac{|\epsilon^{'}(\omega)|}{\epsilon_{d}\delta_{GNDs}}-1\right]=\frac{\epsilon_{0}t\pi (1-\epsilon_{d}\delta_{GNDs})(\hbar^{2}\omega^{2}+\hbar^{2}\gamma_{G})-e^{2}E_{F}}{\pi\delta_{GNDs}\epsilon_{d}\epsilon_{0}t(\hbar^{2}\omega^{2}+\hbar^{2}\gamma_{G})}
\label{A3}
\end{equation}
Thus, the conductivity of the GNDs lattice can be written as:

\begin{align}
\sigma_{GNDs}=&&it \epsilon_{0}\omega +\frac{it(1+\pi/2)\epsilon_{0}\omega}{\left(\frac{1}{\epsilon_{d}}-\frac{e^{2}E_{F}}{\pi \epsilon_{0}\hbar^{2}t\epsilon_{d}(\omega^{2}+\gamma^{2})}+1\right)}
-\frac{\frac{i(1+\pi/2)\omega e^{2}E_{F}}{\pi\hbar^{2}(\omega^{2}+\gamma_{G}^{2})}}{\left(\frac{1}{\epsilon_{d}}-\frac{e^{2}E_{F}}{\pi \epsilon_{0} \hbar^{2}\epsilon_{d}t(\omega^{2}+\gamma_{G}^{2})}+1\right)}\nonumber\\&&-\frac{it\pi \epsilon_{0}  \omega}{\sqrt{\left( \frac{2\pi \epsilon_{0}\hbar^{2}t(\omega^{2}+\gamma_{G}^{2})(1-\epsilon_{d}\delta_{G})-e^{2}E_{F}}{\pi\epsilon_{0} \epsilon_{d} t\delta_{G}\hbar^{2}(\omega^{2}+\gamma_{G}^{2})}-\frac{i2\gamma_{G}}{\omega} \right)\left(\frac{1}{\epsilon_{d}}-\frac{e^{2}E_{F}}{\pi \epsilon_{0} \hbar^{2} \epsilon_{d}t(\omega^{2}+\gamma_{G}^{2})}+1\right)}}\nonumber\\&& +\frac{ \frac{i\omega e^{2}E_{F}}{\pi\hbar^{2}(\omega^{2}+\gamma_{G}^{2})}}{\sqrt{\left( \frac{2\pi \epsilon_{0}\hbar^{2}t(\omega^{2}+\gamma_{G}^{2})(1-\epsilon_{d}\delta_{G})-e^{2}E_{F}}{\pi \epsilon_{0}\epsilon_{d} \delta_{G}\hbar^{2}t(\omega^{2}+\gamma_{G}^{2})}-\frac{i2\gamma_{G}}{\omega} \right)\left(\frac{1}{\epsilon_{d}}-\frac{e^{2}E_{F}}{\pi \epsilon_{0}\hbar^{2}  \epsilon_{d}t(\omega^{2}+\gamma_{G}^{2})}+1\right)}}
\label{A4}
\end{align}
To simplify the calculations of the frequency derivative of $\sigma_{GNDs}$ put:
\begin{equation}
A(\omega_{u})=\left( \frac{2t\pi \epsilon_{0}\hbar^{2}(\omega_{u}^{2}+\gamma_{G}^{2})(1-\epsilon_{d}\delta_{G})-e^{2}E_{F}}{\pi\epsilon_{0}t \hbar^{2}\epsilon_{d} \delta_{G}(\omega_{u}^{2}+\gamma_{G}^{2})}-\frac{i2\gamma_{G}}{\omega_{u}} \right)\nonumber
\end{equation}
This yields:
\begin{equation}
A(\omega_{u})=\left( \frac{2\pi \epsilon_{0}\hbar^{2}t\omega_{u} (\omega_{u}^{2}+\gamma_{G}^{2})(1-\epsilon_{d}\delta_{G})-e^{2}E_{F}\omega_{u}-2i\epsilon_{0}\hbar^{2}\epsilon_{d}\pi \delta_{G} t\gamma_{G}(\omega_{u}^{2}+\gamma_{G}^{2})}{\pi \epsilon_{0}\epsilon_{d} \eta_{g}\hbar^{2}t\omega_{u}(\omega_{u}^{2}+\gamma_{G}^{2})} \right)
\label{A5}
\end{equation}
and:
\begin{equation}
B(\omega_{u})=\left(\frac{1}{\epsilon_{d}}-\frac{e^{2}E_{F}}{\pi \epsilon_{0}\hbar^{2}  \epsilon_{d} t (\omega_{u}^{2}+\gamma_{G}^{2})}+1\right)\nonumber
\end{equation}
which gives:
\begin{equation}
B(\omega_{u})=\frac{-\epsilon_{d}e^{2}E_{F}+\pi \epsilon_{0} (1+\epsilon_{d})t\epsilon_{d}\hbar^{2}(\omega_{u}^{2}+\gamma_{G}^{2})}{\pi \epsilon_{0} \hbar^{2} \epsilon_{d}^{2} t (\omega_{u}^{2}+\gamma_{G}^{2})}
\label{A6}
\end{equation}
Therefore:
\begin{eqnarray}
\frac{dA}{d\omega_{u}}=&&\frac{[2\pi \epsilon_{0}\hbar^{2} t(3\omega_{u}^{2}+\gamma_{G}^{2})(1-\epsilon_{d}\delta_{G})-e^{2}E_{F}-4i\pi \epsilon_{0} \epsilon_{d}t\hbar^{2}\omega_{u} \gamma_{G}][\pi \epsilon_{0} \epsilon_{d}t\delta_{G}\omega_{u} \hbar^{2}(\omega_{u}^{2}+\gamma_{G}^{2})]}{\pi \epsilon_{0} \epsilon_{d}t\delta_{G}\omega_{u} \hbar^{2}(\omega_{u}^{2}+\gamma_{G}^{2})]^{2}}\nonumber\\&&-\frac{[2\pi \epsilon_{0}\omega_{u} \hbar^{2}(\omega_{u}^{2}+\gamma_{G}^{2})(1-\epsilon_{d}\delta_{G})-e^{2}E_{F}\omega_{u}-2i\epsilon_{0}\epsilon_{d}\pi \delta_{G} t\gamma_{G}\hbar^{2}(\omega_{u}^{2}+\gamma_{G}^{2})]}{[\pi \epsilon_{0} \epsilon_{d}t\delta_{G}\omega_{u} \hbar^{2}(\omega_{u}^{2}+\gamma_{G}^{2})]^{2}}\nonumber\\&&\times [\pi \epsilon_{0}\hbar^{2}\epsilon_{d}\delta_{G}t(3\omega_{u}^{2}+\gamma_{G}^{2})]
\label{A7}
\end{eqnarray}
and;
\begin{eqnarray}
\frac{dB}{d\omega_{u}}=&&\frac{[2\pi \hbar^{2}\epsilon_{0}\epsilon_{d}t(1+\epsilon_{d})\omega_{u}][\pi \epsilon_{0}  \epsilon_{d}^{2} t \hbar^{2}(\omega_{u}^{2}+\gamma_{G}^{2})]}{[\pi \epsilon_{0}  \epsilon_{d}^{2} t \hbar^{2}(\omega_{u}^{2}+\gamma_{G}^{2})]^{2}}\nonumber\\&&-\frac{[-\epsilon_{d}e^{2}E_{F}+\pi \epsilon_{0} (1+\epsilon_{d})t\epsilon_{d}\hbar^{2}(\omega_{u}^{2}+\gamma_{G}^{2})][2\pi \epsilon_{0}\hbar^{2}\epsilon_{d}^{2}\omega_{u}]}{[\pi \epsilon_{0}  \epsilon_{d}^{2} t \hbar^{2}(\omega_{u}^{2}+\gamma_{G}^{2})]^{2}}=C(\omega_{u})
\label{A8}
\end{eqnarray}
Thus, the frequency derivative in the denominator of Eq.~\ref{A4} can be given as:
\begin{equation}
\frac{d}{d\omega_{u}}[A(\omega_{u})B(\omega_{u})]^{1/2}=\frac{1}{2}[A(\omega_{u})B(\omega_{u})]^{-1/2}\left[A(\omega_{u})\frac{dB}{d\omega_{u}} +B(\omega_{u}) \frac{dA}{d\omega_{u}}\right]=D(\omega_{u})
\label{A9}
\end{equation}
Substituting from eqs.~(\ref{A5})-~(\ref{A9}) through the use of Eq.~(\ref{A4}) gives:
\begin{eqnarray}
\frac{d\sigma_{GNDs}}{d\omega_{u}}=&& i\epsilon_{0}t-\frac{i\pi \epsilon_{0}t[(AB)^{1/2}-\omega_{u} D(\omega_{u})]}{AB}+\frac{i\epsilon_{0}t(1+\pi/2)(B-\omega_{u} C(\omega_{u}))}{B^{2}}\nonumber\\&&-\frac{\frac{i(1+\pi/2)e^{2}E_{F}}{\pi \hbar^{2}(\omega_{u}^{2}+\gamma_{G}^{2})}\left[ \frac{(\hbar^{2}\omega_{u}^{2}+\hbar^{2}\gamma_{G}^{2})-2\hbar^{2}\omega_{u}^{2}}{(\hbar^{2}\omega_{u}^{2}+\hbar^{2}\gamma_{G}^{2})}B-\omega_{u} C(\omega_{u})\right]}{B^{2}}\nonumber\\&&+\frac{\frac{ie^{2}E_{F}}{(\hbar^{2}\omega_{u}^{2}+\hbar^{2}\gamma_{G}^{2})}\left[ \frac{(\hbar^{2}\omega_{u}^{2}+\hbar^{2}\gamma_{G}^{2})-2\hbar^{2}\omega_{u}^{2}}{(\hbar^{2}\omega_{u}^{2}+\hbar^{2}\gamma_{G}^{2})}(AB)^{1/2}-\omega_{u} D(\omega_{u}) \right]}{(AB)}=E(\omega_{u})
\label{A10}
\end{eqnarray}
\subsection{The frequency derivative of $\sigma_{MNPs}$}
    If we write the dielectric function of metal as; $\epsilon_{M}=\epsilon_{M}^{'}(1+i\kappa_{M})$, then for $\epsilon_{M}^{'}<0$, $\kappa_{M}<<1$, and $\delta_{M}>>1$, we have \cite{genov2004resonant}:
\begin{equation}
\epsilon_{eff}=2\epsilon_{d}\left[ \left(1+\frac{\kappa_{M} \Delta_{M}}{P_{M}+1}\right)log\left( \frac{P_{M}+1}{\kappa_{M}(\Delta_{M}-i)} \right) -1 \right]
\label{A11}
\end{equation}
where $\kappa_{M}=\gamma_{M}/\omega_{u}$. For $ \kappa_{M}<<1$ and $\omega_{p}>>\omega_{u}$, the parameter $P_{M}$ can be approximated as $(\omega_{p}/\omega_{u})^{2}\epsilon_{d}^{-1}$. Thus, Eq.~\ref{A11} can be rewritten as:
\begin{equation}
\epsilon_{eff}=2\epsilon_{d}\left[ \left(1+\frac{\frac{\omega_{p}^{2}}{\omega_{u}^{2}\delta_{M}\epsilon_{d}}-1}{\frac{\omega_{p}^{2}}{\omega_{u}^{2}\epsilon_{d}}+1}\right)log\left( \frac{\frac{\omega_{p}^{2}}{\omega_{u}^{2}\epsilon_{d}}+1}{\left(\frac{\omega_{p}^{2}}{\omega_{u}^{2}\delta_{M}\epsilon_{d}}-1 \right)-i\gamma_{M}/\omega_{u}}\right)-1 \right]
\label{A12}
\end{equation}
Using the following identity, 
\begin{equation}
log(z)=log(x+iy)=ln\sqrt{x^{2}+y^{2}}+i tan^{-1}\left( \frac{y}{x} \right)
\label{A13}
\end{equation}
we can write Eq.~(\ref{A12}) as:
\begin{eqnarray}
\epsilon_{eff}=&&2\epsilon_{d} \left(1+\frac{\frac{\omega_{p}^{2}}{\omega_{u}^{2}\delta_{M}\epsilon_{d}}-1}{\frac{\omega_{p}^{2}}{\omega_{u}^{2}\epsilon_{d}}+1}\right)\left(ln\sqrt{\frac{\left( \frac{\omega_{p}^{2}}{\omega_{u}^{2}\epsilon_{d}}-1 \right)^{2}\left[  \left(\frac{\omega_{p}^{2}}{\omega_{u}^{2}\delta_{M}\epsilon_{d}}-1 \right)^{2}+\left(\frac{\gamma_{M}}{\omega_{u}}\right)^{2}\right]}{\left(\frac{\omega_{p}^{2}}{\omega_{u}^{2}\delta_{M}\epsilon_{d}}-1 \right)^{2}+\left(\frac{\gamma_{M}}{\omega_{u}}\right)^{2}}}-1\right)\nonumber\\&&
+2i\epsilon_{d} \left(1+\frac{\frac{\omega_{p}^{2}}{\omega_{u}^{2}\delta_{M}\epsilon_{d}}-1}{\frac{\omega_{p}^{2}}{\omega_{u}^{2}\epsilon_{d}}+1}\right) tan^{-1}\left(\frac{\gamma_{M}/\omega_{u}}{\frac{\omega_{p}^{2}}{\omega_{u}^{2}\delta_{M}\epsilon_{d}}-1}\right)
\label{A14}
\end{eqnarray}
The relation between the conductivity and the dielectric function of the metallic lattice leads to:
\begin{equation}
\sigma_{MNPs}=2i\epsilon_{0}\epsilon_{d}\omega_{u}\left(1+\frac{\frac{\omega_{p}^{2}}{\omega_{u}^{2}\delta_{M}\epsilon_{d}}-1}{\frac{\omega_{p}^{2}}{\omega_{u}^{2}\epsilon_{d}}+1}\right)tan^{-1}\left(\frac{\gamma_{M}/\omega_{u}}{\frac{\omega_{p}^{2}}{\omega_{u}^{2}\delta_{M}\epsilon_{d}}-1}\right)
\label{A15}
\end{equation}
Therefore:
\begin{eqnarray}
\frac{d\sigma_{MNPs}}{d\omega_{u}}=&&2i\epsilon_{0}\epsilon_{d}\left[ 1+\frac{\left(\frac{-\omega_{p}^{2}}{\omega_{u}^{2}\delta_{M}\epsilon_{d}}-1\right)\left(\frac{\omega_{p}^{2}}{\omega_{u}^{2}\epsilon_{d}}+1\right)+2\frac{\omega_{p}^{2}}{\epsilon_{d}\omega_{u}^{3}}\left(\frac{\omega_{p}^{2}}{\omega_{u} \delta_{M}\epsilon_{d}}-\omega_{u}\right)}{\left( \frac{\omega_{p}^{2}}{\omega_{u}^{2}\epsilon_{d}}+1 \right)^{2}}\right] tan^{-1}\left(\frac{\gamma_{M}/\omega_{u}}{\frac{\omega_{p}^{2}}{\omega_{u}^{2}\delta_{M}\epsilon_{d}}-1}\right)\nonumber\\&& +2i\epsilon_{0}\epsilon_{d}\left( \omega_{u}+\frac{\frac{\omega_{p}^{2}}{\omega_{u} \delta_{M}\epsilon_{d}}-\omega_{u}}{\frac{\omega_{p}^{2}}{\omega_{u}^{2}\epsilon_{d}}+1} \right)\times \frac{\frac{-\gamma_{M}}{\omega_{u}^{2}}\left(\frac{\omega_{p}^{2}}{\omega_{u}^{2}\delta_{M}\epsilon_{d}}-1 \right)+\frac{2\omega_{p}^{2}\gamma_{M}}{\epsilon_{d}\delta_{M}\omega_{u}^{4}}}{\left[ \frac{\omega_{p}^{2}}{\omega_{u}^{2}\delta_{M}\epsilon_{d}}-1\right]^{2}\left[  1+\frac{\left( \frac{\gamma_{M}}{\omega_{u}} \right)^{2}}{\left(\frac{\omega_{p}^{2}}{\omega_{u}^{2}\delta_{M}\epsilon_{d}}-1 \right)^{2}} \right]}=F(\omega_{u})
\label{A16}
\end{eqnarray}

\end{appendix}
\bibliographystyle{spphys}  

\bibliography{Ref}  
\end{document}